\documentclass{article}

\usepackage{arxiv}

\usepackage[utf8]{inputenc} 
\usepackage[T1]{fontenc}    
\usepackage{hyperref}       
\usepackage{url}            
\usepackage{booktabs}       
\usepackage{amsfonts}       
\usepackage{nicefrac}       
\usepackage{microtype}      
\usepackage{lipsum}		
\usepackage{graphicx}
\usepackage[numbers, square]{natbib}
\usepackage{doi}
\usepackage{amsmath}
\usepackage{siunitx}

\title{Revisiting PSF Models: Unifying Framework and High-Performance Implementation}

\author{Yan Liu\textsuperscript{1}
\And
Vasiliki Stergiopoulou\textsuperscript{2,3}
\And
Jonathan Chuah\textsuperscript{1}
\AND
Eric Bezzam\textsuperscript{2}
\And
Gert-Jan Both\textsuperscript{4}
\And
Michael Unser\textsuperscript{1}
\AND
Daniel Sage\textsuperscript{1}
\And
Jonathan Dong\textsuperscript{1}\thanks{Email: jonathan.dong@epfl.ch}
\AND
\\
\textsuperscript{1}Biomedical Imaging Group, École Polytechnique Fédérale de Lausanne, Lausanne, Switzerland \\
\textsuperscript{2}LCAV, École Polytechnique Fédérale de Lausanne, Lausanne, Switzerland \\
\textsuperscript{3}GALATEA, École Polytechnique Fédérale de Lausanne, Lausanne, Switzerland\\
\textsuperscript{4}HHMI Janelia Research Campus, Ashburn VAUSA}

\newcommand{\efarvec}{\mathbf{e}_\infty}
\newcommand{\efar}{e_\infty}
\newcommand{\eincvec}{\mathbf{e}_\mathrm{inc}}
\newcommand{\einc}{e_\mathrm{inc}}
\newcommand{\eincx}{e_\mathrm{inc}^x}
\newcommand{\eincy}{e_\mathrm{inc}^y}
\newcommand{\einca}{e_\mathrm{inc}^a}

\renewcommand{\headeright}{}
\renewcommand{\undertitle}{}
\renewcommand{\shorttitle}{}

\hypersetup{
pdftitle={Revisiting PSF models},
pdfsubject={},
pdfauthor={Y. Liu, V. Stergiopoulou, J. Chuah, E. Bezzam, G.-J. Both, M. Unser, D. Sage, J. Dong},
pdfkeywords={point spread function, localization microscopy, vectorial field propagation, open-source library},
}

\begin{document}
\maketitle

\begin{abstract}
    Localization microscopy often relies on detailed models of point-spread functions.
    For applications such as deconvolution or PSF engineering, accurate models for light propagation in imaging systems with a high numerical aperture are required.
    Different models have been proposed based on 2D Fourier transforms or 1D Bessel integrals.
    The most precise ones combine a vectorial description of the electric field and accurate aberration models.
    However, it may be unclear which model to choose as there is no comprehensive comparison between the Fourier and Bessel approaches yet.
    Moreover, many existing libraries are written in Java (e.g., our previous PSF generator software) or MATLAB, which hinders their integration into deep learning algorithms.
    In this work, we start from the original Richards-Wolf integral and revisit both approaches in a systematic way.
    We present a unifying framework in which we prove the equivalence between the Fourier and Bessel strategies and detail a variety of correction factors applicable to both of them.
    Then, we provide a high-performance implementation of our theoretical framework in the form of an open-source library that is built on top of PyTorch, a popular library for deep learning.
    It enables us to benchmark the accuracy and computational speed of different models and allows for an in-depth comparison of the existing models for the first time.
    We show that the Bessel strategy is optimal for axisymmetric beams, while the Fourier approach can be applied to more general scenarios.
    Our work enables the efficient computation of a point-spread function on CPU or GPU, which can then be included in simulation and optimization pipelines. 
\end{abstract}


\renewcommand\thefootnote{}
\footnotetext{\textbf{Abbreviations:} PSF, point-spread function; NA, numerical aperture; FFT, fast Fourier transform; CPU, central processing unit; GPU, graphics processing unit; SMLM, single-molecule localization microscopy.}

\renewcommand\thefootnote{\fnsymbol{footnote}}
\setcounter{footnote}{1}

\section{Introduction}\label{sec1}

The point-spread function (PSF), also referred to as the impulse response, is a fundamental concept that encapsulates key features of an optical microscope.
Its monitoring and assessment have been longstanding routines in microscopy, supported by the development of specific tools \cite{matthews2010metroloj, theer2014psfj, miora2024calculating} and the creation of a consortium to standardize best practices \cite{nelson2022monitoring}.
A detailed characterization of the PSF is essential to the design of computational imaging processes such as single-molecule localization microscopy \cite{lelek2021single} and super-resolution microscopy, including 3D deconvolution microscopy \cite{sibarita2005deconvolution, sage2017deconvolutionlab2}, structured illumination microscopy (SR-SIM) \cite{heintzmann1999laterally, gustafsson2000surpassing}, fluctuation microscopy \cite{dertinger2009fast,stergiopoulou_BioIm}, stimulated emission-depletion microscopy (STED) \cite{hell2007far}, and MINFLUX \cite{balzarotti2017nanometer}.
The PSF is at the heart of these localization techniques and helps one to achieve a spatial resolution beyond the optical diffraction limit.

The PSF can be measured experimentally from a z-stack of tiny fluorescent beads \cite{sibarita2005deconvolution, marin2021pymevisualize}. 
However, a theoretical PSF is often required. 
Here, the challenge lies in the accurate model of light propagation for imaging systems with high numerical aperture (NA).
Detailed models have been proposed \cite{richards1959electromagnetic, leutenegger2006fast, aguet2009super, Novotny_Hecht_2012} to (a) go beyond simple paraxial approximations, (b) take into account the vectorial nature of the electric field, and (c) include various aberration factors such as the Gibson-Lanni aberrations due to refraction at planar interfaces \cite{gibson1991experimental}.
Several software packages have been developed to generate precise theoretical 3D PSFs that include various features and aberrations.
Among these, the Huygens PSF software\footnote{\url{https://svi.nl/}} is well-known in the microscopy community, and our Java-based ImageJ/FIJI plugin, PSF Generator \cite{kirshner20133}, has also been widely used in the past. 
Other open-source alternatives include MATLAB-based solutions \cite{nasse2010realistic,miora2024calculating, schneider2024interactive} as well as Python-based tools \cite{caprile2022pyfocus, prigent2023spitfir}. 

Computational PSF models have often been applied in fluorescence microscopy.
For example, the fitting of a theoretical PSF to a measured one can improve accuracy \cite{kirshner20133, li2018real}. 
Detailed models enable the quantification of the 3D uncertainty \cite{dong2021fundamental}, while the engineering of PSFs may take full advantage of parametric models \cite{shechtman2014optimal, opatovski2024depth, Liu:25}. 
They are a key component of high-quality single-molecule localization microscopy (SMLM) images \cite{sinko2014teststorm, sage2015quantitative}, including single-particle tracking \cite{wieser2008tracking, shen2017single} and virtual SMLM microscopes \cite{griffie2020virtual, bourgeois2023single}.

Existing approaches to the computation of the PSF can be categorized into two classes of models: those based on 2D Fourier transforms \cite{leutenegger2006fast, miora2024calculating} or 1D integrals of Bessel functions \cite{richards1959electromagnetic, torok1997electromagnetic, aguet2009super, Novotny_Hecht_2012}.
These two classes rely on different assumptions (with Bessel models being more restrictive) and bring different computational trade-offs.
Most works focus on one of the two approaches, and the relationship between them remains unclear.
For example, the Gibson-Lanni aberrations \cite{gibson1991experimental} or the apodization factor \cite{richards1959electromagnetic} are only applied in the Bessel case, although they could be generalized to both.
To the best of our knowledge, no systematic benchmark of the two strategies has been published, either in terms of accuracy or computational speed.

In this work, we propose a unifying framework for PSF models in which we show the equivalence between the Fourier and the Bessel approaches as being two separate parameterizations of the same propagation integral.
This enables us to generalize diverse correction factors and apply them to both models.
Eventually, the choice is simplified to scalar versus vectorial models, optionally with additional correction factors. 
We distribute an open-source PyTorch-based library called \texttt{psf-generator}. It inherits all the functionalities of PyTorch and allows for a seamless integration with modern learning-based algorithms.
We benchmark the two classes of models on CPU and GPU and show that the Bessel model is advantageous for axisymmetric PSFs while the Fourier one is more general and thus more suitable for applications such as PSF engineering.
To support the open-source community and facilitate the adoption of high-NA PSF models, we also provide a graphical interface in the napari ecosystem \cite{napari} and integrate our work in the chromatix optical simulation framework \cite{deb2025chromatix}.

\section{Background}
Electromagnetic waves in optical systems are fundamentally described by the Helmholtz wave equations, yet solving these equations in full generality is computationally intractable.
Various approximations and integral formulations have thus been developed.
The cornerstone for precise PSF models in high-NA systems is the Richards-Wolf integral \cite{richards1959electromagnetic}, which can be viewed as a vectorial extension of the Debye integral \cite{debye1908lichtdruck}.
The conditions of validity for the Richards-Wolf integral have been thoroughly discussed by \cite{wolf1981conditions}.
While alternative light propagation models exist, the Huygens-Fresnel approach has been shown to be equivalent, to some extent, for PSF calculations \cite{egner1999equivalence}. 

This has been the basis of a line of work for PSF models based on Bessel functions, which we will later refer to as the spherical parametrization of the Richards-Wolf integral.
This approach includes simpler formulations like the Kirchhoff model and more sophisticated vectorial representations \cite{aguet2009super, Novotny_Hecht_2012}.
These models have progressively been refined by incorporating various correction factors to account for additional physical processes: the Gibson-Lanni model for spherical aberrations due to refractive index mismatch \cite{gibson1991experimental} (later generalized in \cite{torok1995electromagnetic,torok1997electromagnetic}), apodization factors for energy conservation \cite{richards1959electromagnetic}, and Fresnel transmission coefficients for accurate interface modeling \cite{aguet2009super}. Models using this spherical parametrization have been implemented in various software libraries in Java \cite{kirshner20133} and Python \cite{caprile2022pyfocus}, which makes it widely accessible to researchers, albeit with some limitations in computational efficiency and integration with deep-learning frameworks.

Another line of work on PSF modeling is based on Fourier transforms, both in scalar \cite{goodman2005introduction} and vectorial \cite{leutenegger2006fast} formulations.
These models are based on a Cartesian parametrization of the underlying Richards-Wolf integral and they represent a more general counterpart of the spherical parametrization.
Recently, these high-NA Fourier models have been implemented in MATLAB \cite{miora2024calculating} and Tensorflow as part of a PSF fitting library \cite{liu2024universal}.
Adequate sampling of the Fourier transform is crucial for obtaining high-resolution PSFs and avoiding aliasing.
A common trick based on the chirp Z transform is usually implemented to achieve it \cite{leutenegger2006fast,miora2024calculating,Liu:25}.

\section{Theory}\label{sec:theory}

\begin{figure}
    \centering
    \includegraphics[width=0.5\linewidth]{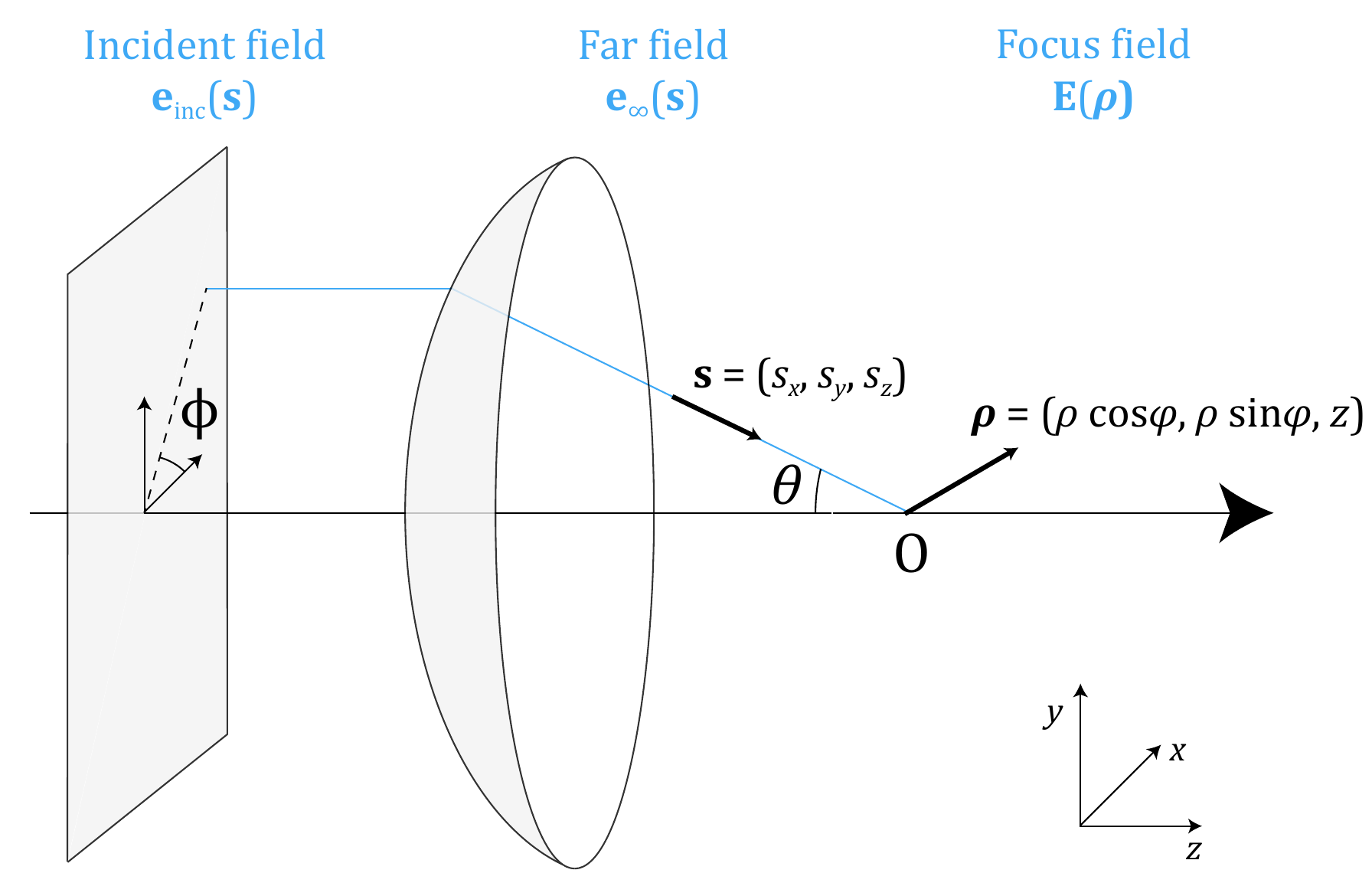}
    \caption{Geometry of the focusing of optical fields.
    An incident field $\eincvec$ is transformed by a focusing element into a converging spherical wave $\efarvec$.
    These fields are parameterized by a unit vector $\mathbf{s}$, either in Cartesian coordinates $(s_x, s_y, s_z)$ or spherical coordinates $(\theta, \phi)$.
    The focus field $\mathbf{E}$ is parameterized by $\boldsymbol{\rho} = (x, y, z)$, rewritten here to introduce cylindrical coordinates $\rho$ and $\phi$.
    }
    \label{fig: 1 geometry}
\end{figure}

\subsection{The Richards-Wolf Model}

As depicted in Figure \ref{fig: 1 geometry}, one obtains the PSF by computing the propagation of light after going through a focusing element, typically a microscope objective or lens.
The incident field $\eincvec(\mathbf{s})$, also called the pupil function, is represented by a disk with a maximal cutoff angle defined by the NA of the imaging system.
The focusing element transforms the planar incident field into a spherical wave, $\efarvec(\mathbf{s})$, evaluated on the Gaussian reference sphere.
This corresponds to an ensemble of far fields that propagate with direction $\mathbf{s}$ and converge to the focal point $O$.
Our goal is to compute the focused electric field $\mathbf{E}(\boldsymbol{\rho})$ near $O$.
Due to the reciprocity of light propagation, this model can also be extended to the emission of a point source to the back focal plane of a microscope objective.

The model introduced by \cite{richards1959electromagnetic} is the starting point that allows us to derive all the precise PSF models described in previous works.
The focal field is given by a sum of plane waves with direction $\mathbf{s} = (s_x,s_y,s_z)$, as expressed by
\begin{equation}
    \mathbf{E}(\boldsymbol{\rho}) 
    = -\frac{\mathrm{i} fk}{2\pi}\iint\limits_{\Omega}\efarvec(\mathbf{s}) \exp{\left(\mathrm{i}k\mathbf{s}\cdot\boldsymbol{\rho}\right)} \mathrm{d}\Omega,
    \label{eq: initial vectorial}
\end{equation}
where $\boldsymbol{\rho} = (x,y,z) = (\rho\cos\varphi, \rho\sin\varphi,z)$ is the position vector in the focal region of the lens,
$\mathbf{s} = (s_x,s_y,s_z) = (\sin\theta \cos\phi, \sin\theta \sin\phi, \cos\theta)$ is a unit vector that describes the direction of an incoming ray,
$f$ is the focal length of the lens,
$k = \frac{2\pi n}{\lambda}$ is the wavenumber,
$\lambda$ is the wavelength,
$n$ is the refractive index of the propagation medium, and $\efarvec(\mathbf{s})$ describes the field distribution on the Gaussian reference sphere.
We integrate over the set $\Omega$ of solid angles defined on a region $s_x^2 + s_y^2 \leq s_\mathrm{max}^2$, where $s_\mathrm{max} = \frac{\mathrm{NA}}{n_i}$ is the cut-off determined by the NA and $n_i$ is the refractive index of the immersion medium.
The angle $\theta$ is therefore defined within the immersion medium. We first describe the different classes of models based on this simplified concise equation and postpone to Section \ref{sec:correction-factors} the introduction of correction factors.

\subsection{Scalar Models}
\label{sec: scalar models}

As a first step, it is common to rely on a scalar approximation to simplify calculations, especially in low-NA scenarios.
In this case, the far field is equal to the incident field, so that $\efar(\mathbf{s}) = \einc(\mathbf{s})$. The focal field is then given by
\begin{equation}
    E(\boldsymbol{\rho}) 
    = -\frac{\mathrm{i} fk}{2\pi}\iint\limits_{\Omega}\efar(s_x,s_y) \exp{\left(\mathrm{i}k\mathbf{s}\cdot\boldsymbol{\rho}\right)} \mathrm{d}\Omega.
    \label{eq: initial scalar}
\end{equation}
This expression involves a two-dimensional integral over the pupil disk.
Two specific parameterizations yield the two classes of models described previously.

The Cartesian parameterization utilizes both $s_x$ and $s_y$ coordinates with $\mathrm{d}\Omega = \mathrm{d}s_x \mathrm{d}s_y / s_z$, which results in
\begin{equation}
    \label{eq: cartesian scalar}
    E(\boldsymbol{\rho}) 
    = -\frac{\mathrm{i} fk}{2\pi}\iint\limits_{s_x^2 + s_y^2 \leq s_{\max}^2} \frac{\efar(s_x,s_y)}{s_z} \exp{\left(\mathrm{i}ks_z z\right)} \exp{\left(\mathrm{i} k (s_x x + s_y y)\right)} \mathrm{d}s_x \mathrm{d}s_y.
\end{equation}
In this form, the focused field at a $z$ transverse plane is given by the 2D inverse Fourier transform of $(\efar(s_x,s_y)\exp{\left(\mathrm{i} k s_z z\right)} / s_z)$, where $s_z = \sqrt{1-s_x^2-s_y^2}$.
Thus, the Cartesian parameterization of the Richards-Wolf integral leverages the speed and efficiency of the fast Fourier transform (FFT) algorithm.

Alternatively, the spherical approach parameterizes the problem with two angles $\theta \in [0, \theta_{\max}]$ (the maximum angle $\theta_{\max}$ is determined by the NA) and $\phi \in [0, 2\pi]$, as depicted in Figure \ref{fig: 1 geometry}.
With $\boldsymbol{\rho} = (x,y,z) = (\rho\cos\varphi, \rho\sin\varphi,z)$ and $\mathrm{d}\Omega = \sin\theta \mathrm{d}\theta \mathrm{d}\phi$, the field in the focal region can be rewritten as
\begin{align}
    \label{eq: polar scalar}
    E(\boldsymbol{\rho}) 
    = -\frac{\mathrm{i} fk}{2\pi} &\int_0^{\theta_{\max}} \mathrm{d}\theta 
    \int_0^{2\pi} \mathrm{d}\phi \, \efar(\theta,\phi) 
    \nonumber \\
    &\exp{\left(\mathrm{i}k\rho\sin\theta\cos(\phi-\varphi)\right)} \exp{\left(\mathrm{i}kz\cos\theta\right)} \sin\theta.
\end{align}
Equation \eqref{eq: polar scalar} can be further simplified if one assumes that the pupil function is axisymmetric (rotational invariant), in the sense that $\efar(\theta, \phi) = \efar(\theta)$.
In this case, the integral over $\phi$ in \eqref{eq: polar scalar} can be computed explicitly using the Bessel function $J_0$\footnote{The spherical parameterization often uses the following identities, where $J_n$ is the Bessel function of $n$th-order of the first kind:\begin{align*}
    &\int_{0}^{2\pi} \cos(n\phi) \exp{\left(\mathrm{i} x \cos(\phi-\varphi)\right)}\mathrm{d}\phi = 2\pi(\mathrm{i})^n J_n(x)\cos(n\varphi)\\
    &\int_{0}^{2\pi} \sin(n\phi) \exp{\left(\mathrm{i} x \cos(\phi-\varphi)\right)}\mathrm{d}\phi = 2\pi(\mathrm{i})^n J_n(x)\sin(n\varphi)
\end{align*}
}, which leads to:
\begin{equation}
    \label{eq: polar scalar final}
    E(\boldsymbol{\rho}) 
    = -\mathrm{i} fk \int_0^{\theta_{\max}} \mathrm{d}\theta\, \efar(\theta)\, J_0(k \rho \sin \theta) \exp{\left(\mathrm{i}kz\cos\theta\right)} \sin\theta.
\end{equation}

Defocus is included in these models via the defocus phase factor $\exp{\left\{\mathrm{i}k s_z z\right\}} = \exp{\left(\mathrm{i}kz \cos\theta\right)}$ where $z$ is the defocus distance.
This expression, also known as angular spectrum propagation \cite{goodman2005introduction}, accurately models the propagation of an electric field in a homogeneous medium. 


\subsection{Vectorial Models}
The electric field is a vectorial quantity. Consequently, vectorial propagation models are necessary to accurately account for the propagation and crosstalk between the components of the vector field.
Precise vectorial models are crucial for high-NA systems, where the need to consider high angles arises.

In the vectorial model, the dependence on the incident field $\eincvec(\mathbf{s})$ of the far field $\efarvec(\mathbf{s})$ now requires us to perform the basis change from a cylindrical to a spherical coordinate system, as in
\begin{align}
    \label{eq: field Gaussian sphere}
    \efarvec(\theta,\phi) = & \left[ \begin{array}{c}
    q_s(1-\cos2\phi) + q_p(1+\cos2\phi)\cos\theta
    \\
    (-q_s+q_p\cos\theta)\sin2\phi \\
    -2 q_p \cos\phi \sin\theta 
\end{array} \right] \frac{\eincx(\theta,\phi)}{2} \nonumber \\  + &
\left[ \begin{array}{c}
    (-q_s+q_p\cos\theta)\sin2\phi  \\
    q_s (1+\cos2\phi) + q_p (1-\cos2\phi) \cos\theta \\
    - 2 q_p \sin\phi \sin\theta 
\end{array} \right] \frac{\eincy(\theta,\phi)}{2}, 
\end{align}
where $\eincvec = [\eincx, \eincy, 0]$. 
The so-called Fresnel transmission coefficients $q_s$ and $q_p$ have been introduced to account for the partial reflections at interfaces, which depend on the polarization state and incidence angle.
For each polarization, the coefficients correspond to the product of all transmission coefficients for each interface from medium $m$ to $m+1$, with
\begin{align}
    &q_s^{m} = \frac{2 n_m \cos{\theta_m}}{n_m\cos\theta_m+n_{m+1}\cos{\theta_{m+1}}} \nonumber\\
    &q_p^{m} = \frac{2 n_m \cos{\theta_m}}{n_{m+1}\cos\theta_m+n_{m}\cos{\theta_{m+1}}}.
\end{align}
It is worth noting that in the low-NA limit ($\theta = 0$) with $q_s = q_p = 1$, \eqref{eq: field Gaussian sphere} simplifies to $\efarvec(\theta,\phi) = \eincvec(\theta, \phi)$. This explains why the scalar model did not account for this geometric change of variables.

The Cartesian parameterization of the vectorial model consists of the integral
\begin{equation}
    \label{eq: cartesian vectorial}
    \mathbf{E}(\boldsymbol{\rho}) 
    = -\frac{\mathrm{i} fk}{2\pi}\iint\limits_{s_x^2+s_y^2 \leq s_{\max}^2}\frac{\efarvec(s_x, s_y)}{s_z}\exp{\left(\mathrm{i}k s_z z\right)}\exp{\left(\mathrm{i}k(s_x x + s_y y)\right)} \mathrm{d}s_x \mathrm{d}s_y,
\end{equation}
which essentially boils down to the computation of the inverse Fourier transform of $(\efarvec(s_x, s_y) \exp{\left(\mathrm{i}k s_z z\right)}/s_z)$, similar to the scalar case. 

With the help of coordinate transformations similar to the scalar case, we derive the spherical parameterization of the field in the focal region as
\begin{align}
    \label{eq: polar vectorial}
    \mathbf{E}(\boldsymbol{\rho}) = -\frac{\mathrm{i} fk}{2\pi}&\int_0^{\theta_{\max}} \mathrm{d}\theta 
    \int_0^{2\pi} \mathrm{d}\phi\,\efarvec(\theta,\phi) 
    \nonumber \\ 
    &\exp{\left(\mathrm{i}k\rho\sin\theta\cos(\phi-\varphi)\right)} \exp{\left(\mathrm{i}kz\cos\theta\right)} \sin\theta.
\end{align}
Inserting \eqref{eq: field Gaussian sphere} into \eqref{eq: polar vectorial} and using the axisymmetric assumption of the incident field, we obtain a simplified expression of the focal field as
\begin{equation}
    \mathbf{E}(\boldsymbol{\rho}) = - \frac{\mathrm{i} fk}{2} \left[ \begin{array}{c}
    [I_{0}^x - I_{2}^x\cos2\varphi] - I_{2}^y\sin2\varphi\\
    - I_{2}^x\sin2\varphi + [I_{0}^y + I_{2}^y\cos2\varphi]\\
    -2\mathrm{i}I_{1}^x\cos\varphi  -2\mathrm{i}I_{1}^y\sin\varphi
  \end{array} \right],
  \label{eq: polar vectorial final}
\end{equation}
where  
\begin{align}
    \label{eq: I0}
    I_{0}^a (\rho,z) &= \int_0^{\theta_{\max}} \einca(\theta)\sin\theta (\cos\theta+1)\,J_0(k\rho\sin\theta)\exp{\left(\mathrm{i}kz\cos\theta\right)}\mathrm{d}\theta \nonumber\\
    I_{1}^a (\rho,z) &= \int_0^{\theta_{\max}} \einca(\theta)\sin^2\theta\, J_1(k\rho\sin\theta)\exp{\left(\mathrm{i}kz\cos\theta\right)}\mathrm{d}\theta \nonumber\\
    I_{2}^a (\rho,z) &= \int_0^{\theta_{\max}} \einca(\theta)\sin\theta (\cos\theta-1) \, J_2(k\rho\sin\theta)\exp{\left(\mathrm{i}kz\cos\theta\right)}\mathrm{d}\theta
\end{align}
with $a\in\{x,y\}$. 

\subsection{Correction Factors}\label{sec:correction-factors}

\begin{figure}[t]
    \centering
    \includegraphics[width=0.5\columnwidth]{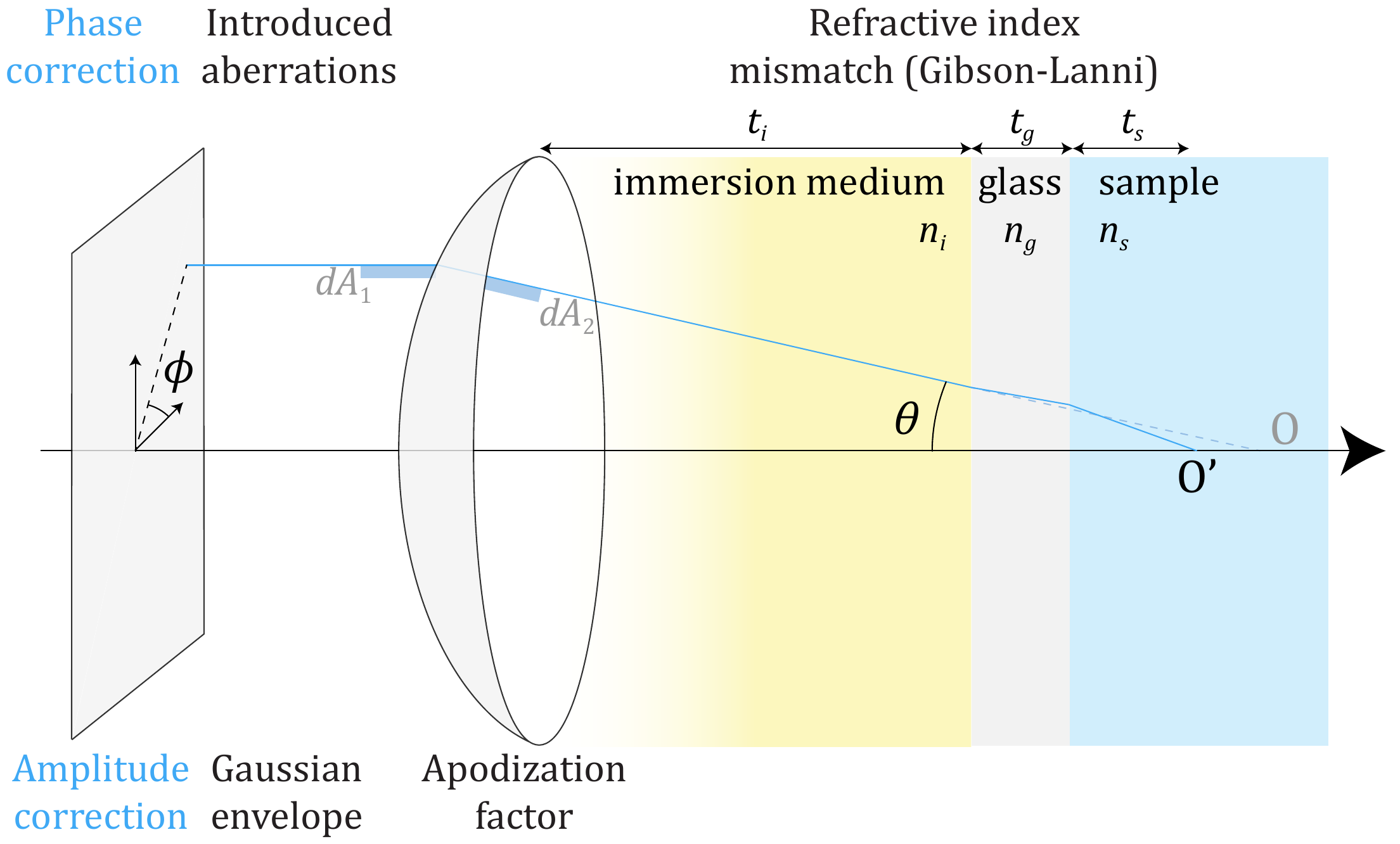}
    \caption{Correction factors and their associated physical origins.
    Top part: phase correction factors, either introduced on the incident field or due to refraction.
    Bottom part: amplitude correction factors that model the incident beam envelope or the apodization factor for energy conservation.}
    \label{fig:correction-factors}
\end{figure}

Precise PSF models commonly take into account several physical effects which may affect the true PSF.
They are added as amplitude factors $a(\mathbf{s})$ and phase factors $W(\mathbf{s})$ in the original integral over solid angles in \eqref{eq: initial vectorial}, leading to
\begin{equation}
    \label{eq: vectorial general with correction}
    \mathbf{E}(\boldsymbol{\rho}) 
    = -\frac{\mathrm{i} fk}{2\pi}\iint\limits_{\Omega} a(\mathbf{s}) \exp{\left(\mathrm{i} W(\mathbf{s})\right)} \efarvec(\mathbf{s}) \exp{\left(\mathrm{i}k\mathbf{s}\cdot\boldsymbol{\rho}\right)} \mathrm{d}\Omega.
\end{equation}
Equation \eqref{eq: vectorial general with correction} enables us to express these correction factors in full generality for both the vectorial and scalar models and for both the Cartesian and spherical parameterizations.
We present a detailed list of these correction factors in Section \ref{sec:ri-mismatch}-\ref{sec:gaussian-envelope} and their graphical description in Figure \ref{fig:correction-factors}. 

\subsubsection{Gibson-Lanni aberrations}\label{sec:ri-mismatch}

Microscopes typically have stratified layers of different refractive indices.
The biological sample is usually aqueous, on top of which we place a coverslip made of glass, and the whole sample is then put in a water or oil immersion medium to increase the numerical aperture.
The microscope objectives are designed to provide aberration-free images in a specific setting with design values for refractive indices and thicknesses of each layer.
Any mismatch introduces spherical aberrations due to refraction at the interface between the layers.
These aberrations can be computed using the formula\footnote{Correction factors will be expressed in spherical coordinates, with $\sin\theta$ being computed in the Cartesian case as $\sin\theta = \frac{NA}{n_i}\sqrt{s_x^2+s_y^2}$.}
\begin{align}
    \label{eq: stratified layers general formula}
    W(\mathbf{s}) =& \frac{2\pi}{\lambda} \left( t_s \sqrt{n_s^2 - n_i^2 \sin^2 \theta} + t_i \sqrt{n_i^2 - n_i^2 \sin^2 \theta} -t_i^* \sqrt{\left.n_i^*\right.^2 - n_i^2 \sin^2 \theta} \right. \nonumber \\
    & \quad \left. + t_g \sqrt{n_g^2 - n_i^2 \sin^2 \theta} - t_g^* \sqrt{\left.n_g^*\right.^2 - n_i^2 \sin^2 \theta}\right),
\end{align}
where $n_s$, $n_i$, $n_g$ are the refractive indices of the sample, immersion medium, and glass, respectively, $t_s$, $t_i$, $t_g$ are the thicknesses of the sample, immersion medium, and glass respectively, and their counterparts with stars are the design conditions. 

In practice, it is challenging to assess the thickness of the immersion medium.
Since this distance is manually tuned to obtain an optimal focus of a point emitter at depth $t_s$ on the camera, this focusing condition gives the relation
\begin{equation}
    \label{eq: gibson lanni t_i}
    t_i = t_s + n_i \left( -\frac{t_s}{n_s} - \frac{t_g}{n_g} + \frac{t_g^*}{n_g^*} + \frac{t_i^*}{n_i^*} \right)
\end{equation}
which can be inserted in \ref{eq: stratified layers general formula}.
This particular expression has first been derived for the spherical scalar case in \cite{gibson1991experimental} and extended to the spherical vectorial case in \cite{torok1995electromagnetic, torok1997electromagnetic}.

\subsubsection{Arbitrary Phase Aberrations}\label{sec:arbitrary}

Refined aberration models can be introduced to describe imperfections in the optical system, for the benefit of PSF engineering or wavefront shaping.
Designed aberrations can be introduced purposefully via a phase mask or a spatial light modulator at the pupil plane. 
The aberrations are often parameterized by Zernike polynomials (a set of orthonormal polynomials defined on the pupil disk) or by a direct fixed phase mask to obtain desired PSFs.
We write it in the most general case as
\begin{equation}
    \label{eq: Zernike aberrations}
    W(\mathbf{s}) = \sum_{k=0}^{K-1} c_k Z_k(\mathbf{s}) + W_0(\mathbf{s}).
\end{equation}
Equation \eqref{eq: Zernike aberrations} is composed of the sum of two terms. The first term is an inner product of the first $K$ Zernike polynomials and their corresponding coefficients $c_k$. The second term can be used to include special cases not covered by the Zernike polynomials (e.g., a vortex phase ramp that leads to a donut PSF as is typically used in STED).

Since the arbitrary phase aberrations described in Equation \eqref{eq: Zernike aberrations} may not necessarily be axisymmetric, they can only be applied to the most general, Cartesian parameterization.

\subsubsection{Apodization Factor}\label{sec:apod}

The apodization factor is an amplitude-correction factor that ensures energy conservation during the change of basis from cylindrical coordinates (incident field $\eincvec$) to spherical coordinates (far field $\efarvec$), which takes prominence especially for high-NA objectives.
Since areas of cross-sections are modified, the field is also rescaled accordingly. Such rescaling ensures that the differential areas $dA_1$ on the plane and $dA_2$ on the sphere, as shown in Figure \ref{fig:correction-factors}, remain consistent under the change of coordinates.
The correcting factor is 
\begin{equation}
    A(\mathbf{s}) = \sqrt{\cos\theta}
\end{equation} 
when going from cylindrical to spherical in the focusing configuration of Figure \ref{fig: 1 geometry}. 

\subsubsection{Gaussian Envelope}\label{sec:gaussian-envelope}

The incident illumination can also depart from a perfect uniform plane wave.
In particular, we often assume a Gaussian envelope and expressed it as
\begin{equation}
    A(\mathbf{s}) = \exp\left(-\frac{\sin^2\theta}{s_\mathrm{env}^2}\right),
\end{equation}
where the constant $s_\mathrm{env}$ determines the size of the envelope.

\section{Implementation}

\subsection{PyTorch library}

We provide a high-performance open-source PyTorch library \texttt{psf-generator}\footnote{\url{https://github.com/Biomedical-Imaging-Group/psf_generator}} to generate 2D and 3D PSFs. 
The library implements the four PSF models described in Section \ref{sec:theory}:
\begin{itemize}
    \item \texttt{ScalarCartesianPropagator} \eqref{eq: cartesian scalar}; 
    \item \texttt{ScalarSphericalPropagator} \eqref{eq: polar scalar final}; 
    \item \texttt{VectorialCartesianPropagator} \eqref{eq: cartesian vectorial}; 
    \item \texttt{VectorialSphericalPropagator} \eqref{eq: polar vectorial final}.
\end{itemize}
Users can choose between these propagator types, as well as configure physical (e.g., numerical aperture, wavelength, and field of view), and numerical parameters (e.g., image dimensions and number of z-planes). 
Our library also allows the users to freely apply any kind of correction factors tailored to their microscope on any chosen propagator as described in Section \ref{sec:arbitrary}. 
The propagators use these parameters to first define the far field (pupil) and propagate it to obtain the focus field (PSF).
Finally, the user can visualize, save, and export the generated PSF using our built-in utility functions. 
Written in PyTorch, the library easily integrates into deep-learning workflows, leveraging native features of PyTorch, such as automatic differentiation. 

In our unifying framework, these models encompass all previously proposed PSF models based on the Richards-Wolf integral.
As discussed in Section \ref{sec: scalar models}, the spherical propagators require the axisymmetric assumption. 
When it is valid, the Cartesian and spherical propagators perform equivalent integral computations and result in the same PSF.
Both are proposed in our library as they differ in terms of computational efficiency and applicability. 

Any tensor corresponding to a field has a shape of the form \texttt{(z, channel, x, y)} to benefit from PyTorch performance optimization, both for data loading and computation parallelization. 
Computation are performed plane-by-plane on the axial \texttt{z} axis, while \texttt{x} and \texttt{y} correspond to the transverse sizes of 2D images. 
Akin to grayscale versus RGB images, the \texttt{channel} dimension is equal to $1$ for scalar models and $3$ in the vectorial case. 
We use the \texttt{ZernikePy} library\footnote{\url{https://pypi.org/project/ZernikePy/}} to generate Zernike polynomials. 

The Cartesian parameterization relies on multiple calls of 2D Fourier transforms with a desired field of view and pixel size. 
To account for the very small physical dimensions associated to pixels in localization microscopy, we have implemented a custom variant of the \texttt{2D FFT} in PyTorch \cite{leutenegger2006fast, miora2024calculating, Liu:25}. 
It enables arbitrary pixel sizes and is based on the chirp Z-transform, at the cost of a convolution and three FFTs.
The computational complexity of a single plane is still $O(n \log n)$, with $n$ the size of the transverse plane.
The size of the Fourier transform is doubled as we only keep the valid convolution range. 

The spherical parameterization relies on the fast and accurate evaluation of multiple one-dimensional integrals that involve the Bessel function of first order $J_0$ over the interval $[0, \theta_{\max}]$, as defined in Eq. \eqref{eq: polar scalar final} for the scalar case.
To maximize speed, we vectorize the computation of the batch of 1D integrals at different defocus distances via \emph{torch.vmap}.
Then, for each 1D integral, we adopt the composite Simpson's rule \cite{stoer_introduction_2002} to benefit from its high accuracy, good numerical stability, and minimal computational overhead.
The computational complexity in this case is $O(n)$, with $n$ the number of integration steps in the interval $[0, \theta_{\max}]$.
As automatic differentiation of Bessel functions is not natively supported by PyTorch as of version 2.3, we also provide a differentiable version of the Bessel functions in our library. 

\subsection{Open-source interoperability}

\begin{figure*}
    \centering
    \includegraphics[width=\linewidth]{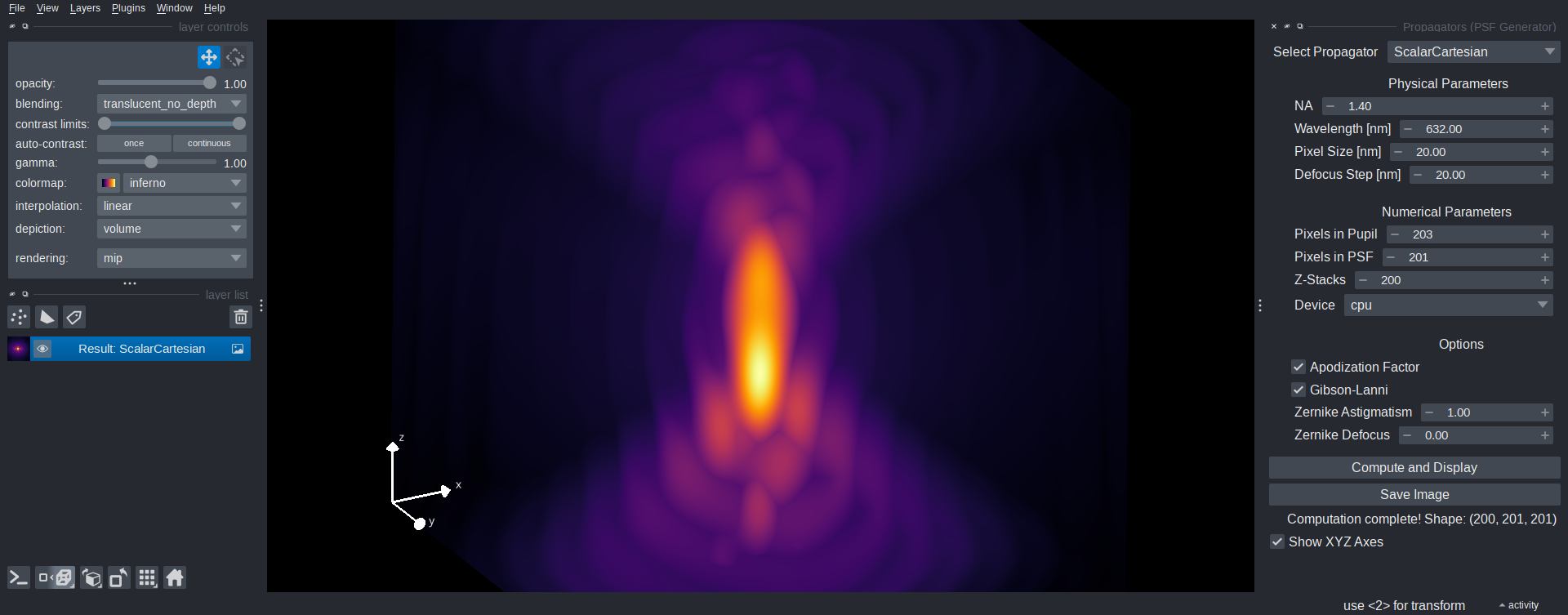}
    \caption{Graphical user interface of our PSF simulation plugin integrated with napari (v0.4.0, 2025-06-17). Left: napari viewer panel enabling interactive 3D navigation and rendering options. Middle: interactive visualization area, displaying here a slice of a high-NA 3D PSF. Right: plugin interface for selecting the propagator model, specifying physical and numerical parameters (e.g., NA, wavelength, Zernike aberrations, CPU or GPU backend), and exporting results.}
    \label{fig:napari-interface}
\end{figure*}

Motivated by FAIR principles for data and code management \cite{barker2022introducing}, we not only open-source our implementations but make them interoperable with other tools. Our goal is to further support the dissemination and adoption of accurate physical models for high-NA light focusing. We present two complementary examples: an intuitive and rapid visualization tool aimed at end users and an integration with a comprehensive optical simulation library intended for researchers. 

\textbf{napari} is an open-source, Python-based image viewer designed for interactive visualization and analysis of multidimensional biological and scientific image data. Built for researchers and developers working with large imaging data from optical microscopy and other imaging modalities, it provides a high-performance interface for exploring, annotating, and processing images. napari is also extensible via plugins, benefiting from the scientific Python ecosystem.

We provide an interface to generate and visualize 3D PSFs using our PyTorch-based framework, as depicted in Figure \ref{fig:napari-interface}. Available as a plugin on the napari hub\footnote{\url{https://www.napari-hub.org/plugins/napari-psfgenerator}}, users can select between the four different propagation models and configure a set of physical and numerical parameters through a user-friendly graphical interface. This facilitates the introduction of high-NA correction factors to all four propagators types. PSFs are computed in real-time and rendered instantly, with support for both CPU and GPU backends. Our plugin allows immediate visual feedback and options to save the resulting volumes. It is designed to provide a code-free solution for the generation, visualization, and export of high-NA PSFs for downstream applications. 
\begin{figure}[tb]
    \centering
    \includegraphics[width=0.65\columnwidth]{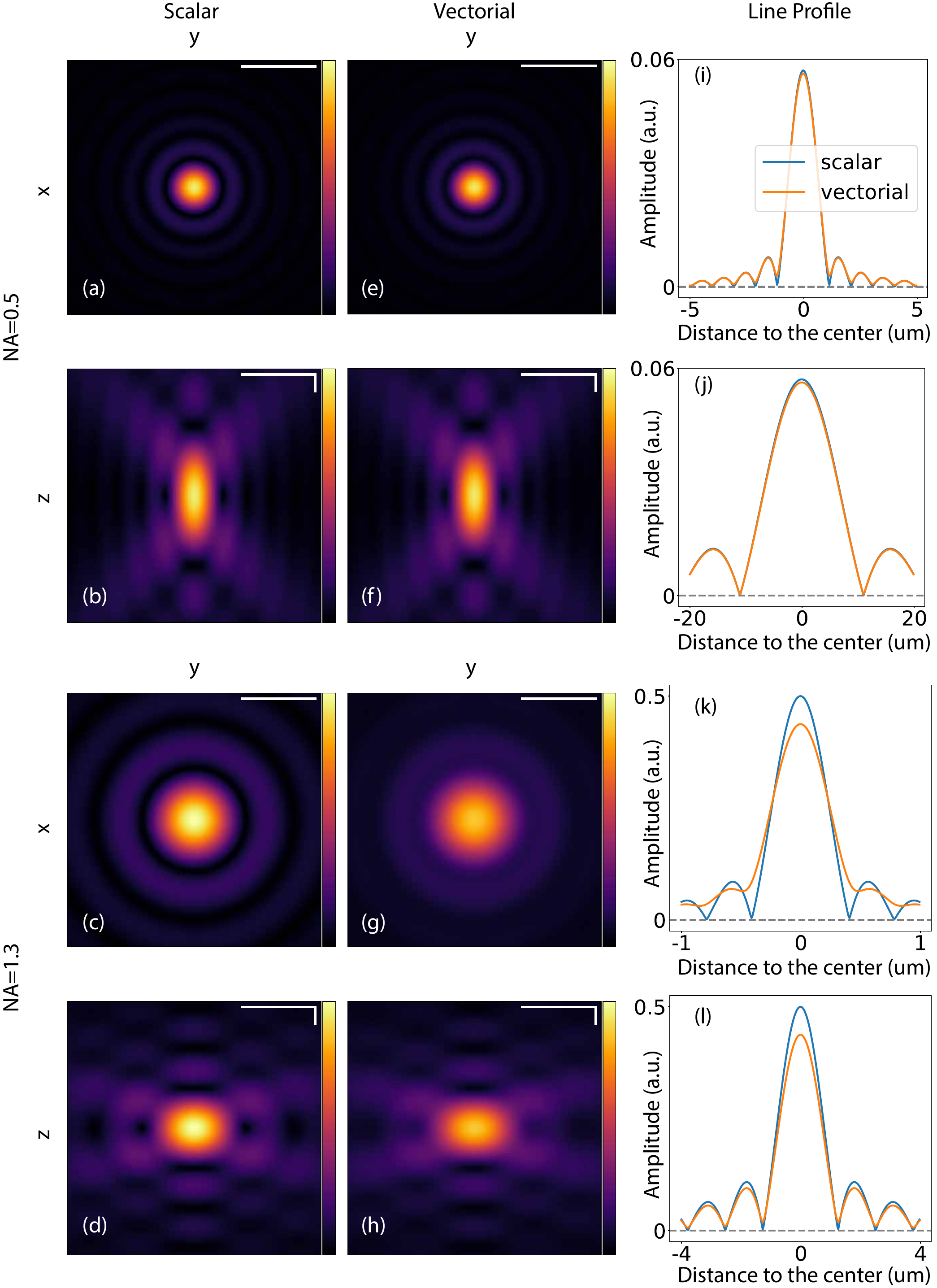}
    \caption{Unaberrated PSFs generated by the scalar ((a)-(d)) and vectorial ((e)-(h)) models, in the case of low-NA (0.5) (first two rows) and high-NA (1.3) (last two rows).
    (a), (e), (c), and (g): Slice of the $x$-$y$ plane.
    (b), (f), (d), and (h): Slice of the $z$-$y$ plane.
    (i)-(l): Intensity profiles along a vertical line through the center of the PSFs.
    The $x$-axis shows the relative distance to the center of the image in micrometers.
    Scale bars represent \qty{3}{\micro\meter} and \qty{0.6}{\micro\meter} in the low- and high- NA cases, respectively.
    For images of $z$-$y$ planes ((b), (f), (d), and (h)), scale bars for the $y$ and $z$ axes are indicated by the horizontal and vertical bars, respectively.
    }
    \label{fig:pure-gaussian-psf}
\end{figure}

\textbf{Chromatix} is an open-source wave optics library capable of simulating a wide range of optical systems such as light field imaging, holography and quantitative phase imaging \cite{deb2025chromatix}. Built on JAX, it supports GPU acceleration and full end-to-end differentiability, making it particularly suitable for large-scale inverse problems and integration with deep learning applications. Compared to PyTorch, JAX offers more seamless support for automatic differentiation and just-in-time (JIT) compilation, which can be advantageous for optimizing physics-based simulations based on wave optics. 

To ensure maximal performance and seamless integration with the rest of the library, the high-NA focusing models have been reimplemented in JAX. We focused on the vectorial Cartesian parametrization, which offers greater generality than the spherical formulation and supports arbitrary pupil fields and aberrations. This functionality is introduced via the \texttt{high\_na\_ff\_lens()} function, which complements the existing \texttt{ff\_lens()} implementation used to model the Fourier transform of low-NA lenses. This addition enables Chromatix to accurately model high-NA microscope systems and facilitates integration into existing simulation pipelines. 

These two examples highlight the importance of actively promoting interoperability within the open-source ecosystem. Nevertheless, all figures and benchmarks presented in this work were produced based on our comprehensive PyTorch-based library, which serves as the foundation of our study.

\section{Results}

\begin{figure}[tb]
    \centering
    \includegraphics[width=0.65\columnwidth]{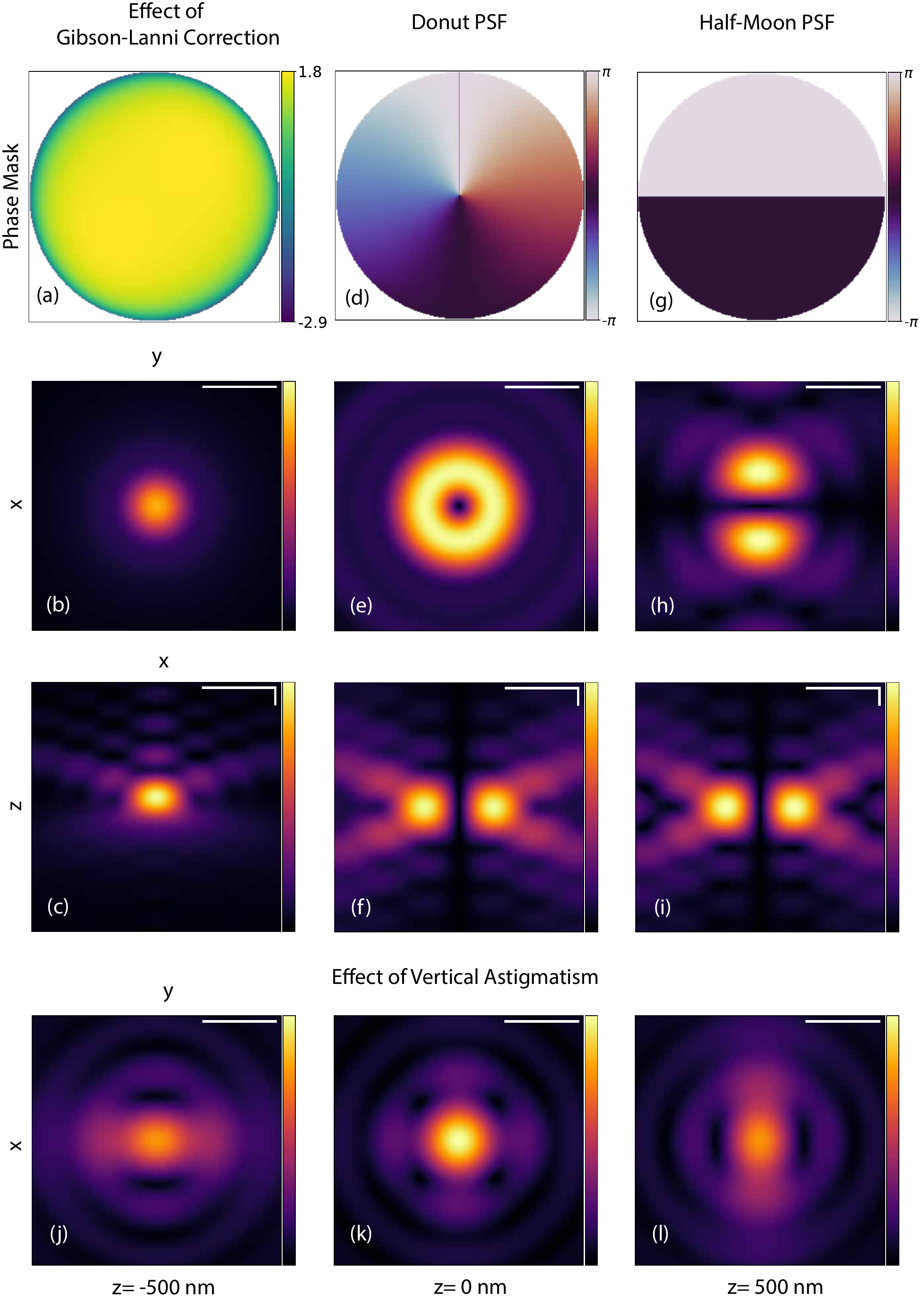}
    \caption{
    PSFs with phase aberrations generated by the vectorial Cartesian propagator at high-NA (1.3). 
    (a), (d), and (g): Introduced phase masks at the pupil plane.
    (b), (e), and (h): Slice of the $x$-$y$ plane at $z=0$.
    (c), (f), and (i): Slice of the $z$-$x$ plane at $y=0$.
    (j)-(l): PSF with vertical astigmatism at three z-planes (in focus (k), \qty{500}{\nano\meter} above (j), and \qty{500}{\nano\meter} below focus (l)).
    All PSFs were generated with circular polarization, except the Half-Moon PSF where x-axis linear polarization is being used. 
    The scale bars in all images represent \qty{0.6}{\micro\meter}.
    }
    \label{fig:other-psfs}
\end{figure}

\subsection{PSF Gallery}

We provide in Figures \ref{fig:pure-gaussian-psf} and \ref{fig:other-psfs} a gallery of PSFs to showcase our PyTorch implementation.
We used a wavelength of 632 nm, a circular polarization for the vectorial models, and we display the amplitude (in arbitrary units) of the beams for better contrast. 
The 2D slices of the same 3D beam share the same dynamic range.
In Figure \ref{fig:pure-gaussian-psf}, we present the focal electric field computed using both the scalar and vectorial models for objectives with different NAs.
In the low-NA case (top two rows), the resulting PSFs from both propagators are similar.
The beams differ more from each other, as expected, in the high-NA case (bottom two rows): the rings are blurred out as the energy is spread into different components of the focus field. 

In Figure \ref{fig:other-psfs}, we present additional PSFs computed with the vectorial propagator in the high-NA setting.
The impact of the Gibson-Lanni correction factor on a beam is shown in Figure \ref{fig:other-psfs} (a)-(c).
The refractive indices and thicknesses of the sample, immersion medium, and glass coverslip are $n_s=1.3$, $n_i=1.5$, $n_g=1.5$, $t_s=$ \qty{1}{\micro\meter}, $t_g=$ \qty{170}{\micro\meter}, while $t_i$ is computed using Eq. \eqref{eq: gibson lanni t_i}.
We observe that the spherical aberration it introduces degrades the quality of the focus.
We also show the donut PSF, which has a vortex phase mask in the pupil plane, (Figure \ref{fig:other-psfs} (d)-(f)) and the half-moon PSF, which has a $\pi$-phase jump in its pupil plane (Figure \ref{fig:other-psfs} (g)-(i)).
Finally, we demonstrate an example with arbitrary phase aberrations using the Zernike polynomials in Figure \ref{fig:other-psfs} (j)-(l). 
Here, some amount of astigmatism is introduced, as is often done to encode defocus information \cite{kao1994tracking}, and we show the evolution of the beam shape along the $z$-axis. 

\subsection{Computational Performance}
\subsubsection{Speed Benchmark}
\begin{figure}[t]
    \centering
    \includegraphics[width=0.5\columnwidth]{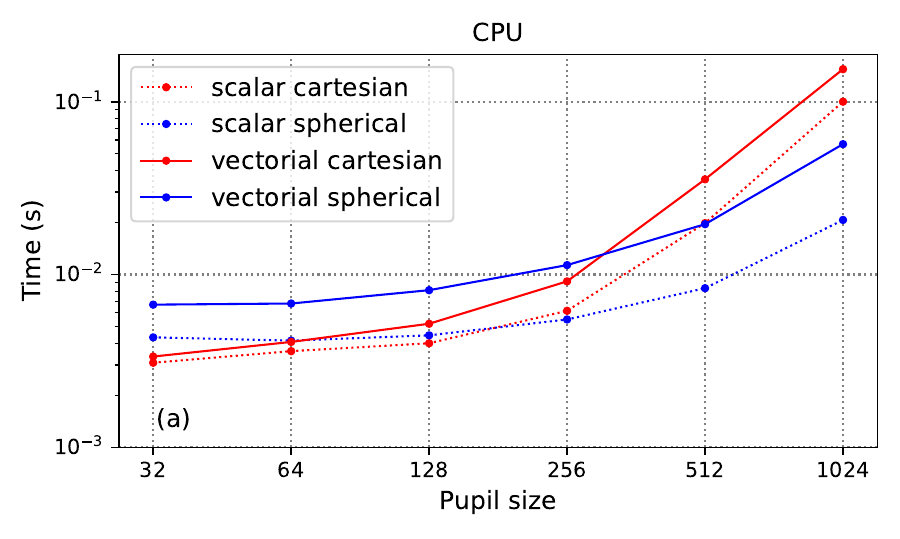}
    \includegraphics[width=0.5\columnwidth]{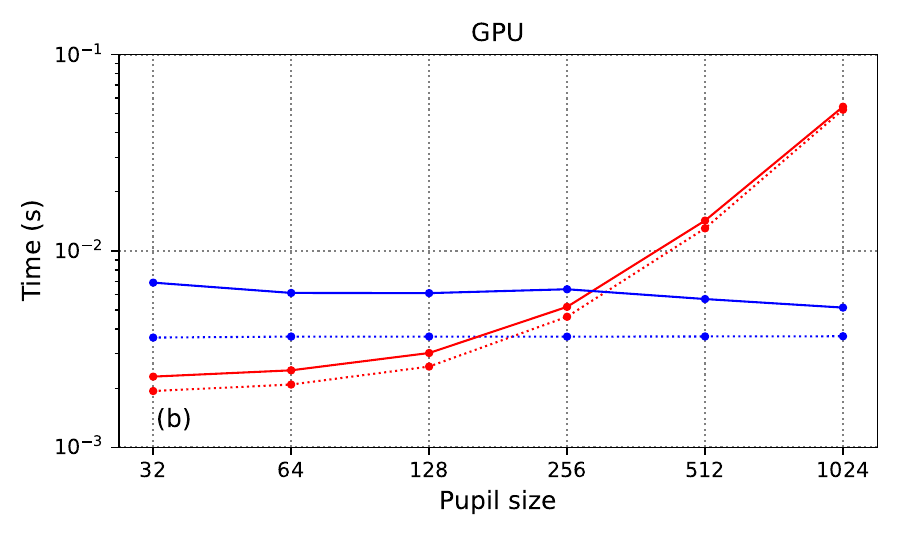}
    \caption{
    Time to generate a $201 \times 201$ PSF in terms of the numerical size of the pupil and the propagator, on CPU (a) and GPU (b). 
    Each data point in the plots is averaged over 10 runs.}
    \label{fig:benchmark-runtime-pupil}
\end{figure}

We benchmark the runtime to compute a single $201 \times 201$ 2D PSF image against the size of the pupil.
The image is captured at focus and all input parameters take default values in our library.
We compare the runtime of all four propagators on CPU and GPU and show the results in Figure \ref{fig:benchmark-runtime-pupil}.
The benchmarking was performed on a machine with an Intel i9-10900X CPU and an NVIDIA GeForce RTX 3090 GPU.

We observe that the runtime of all propagators increases with the size of the pupil on CPU.
The Cartesian propagators (red) are faster than the spherical ones (blue) on small sizes (<512 pixels) but slower on larger sizes.
This illustrates the computational complexity of each method. 
Scalar propagators (dotted) are faster than their vectorial counterparts (solid), by roughly 1.5 times for the Cartesian and 3 times for the spherical cases.
On GPU, Cartesian propagators are faster than on CPU at the same grid size and the curves (red) behave similarly. Spherical propagators, however, exhibit a flat curve (blue), which indicates that they benefit heavily from the GPU parallelization.
The speed improvement between scalar and vectorial propagators is small, especially for the Cartesian case.
Hence, vectorial propagators should be preferred over the scalar ones as the accuracy gain does not come at a high computational cost.
Moreover, Cartesian propagators are recommended if one works with images of small size for both scalar and vectorial cases. For larger sizes, however, spherical propagators should be preferred, especially when a GPU is available.

\subsubsection{Accuracy Benchmark}\label{sec:benchmark-accuracy}
We benchmark the accuracy of the Cartesian and spherical scalar propagators with the analytic Airy disk for asymptotic limit (the Fourier transform of a perfect circular aperture) given by
\begin{equation}
\label{eq:airy-disk-analytic}
    F_{\mathrm{AD}}(\rho) = \frac{2J_1(\rho)}{\rho},
\end{equation}
where $J_1$ is the Bessel function of the first order of the first kind. As the Airy disk is the Fourier transform of a perfect unit modulus circular aperture, an additional factor $s_z = \cos\theta$ is introduced in Eqs. \eqref{eq: cartesian scalar} and \eqref{eq: polar scalar final}. This corresponds to a paraxial approximation. 

The spherical integral is computed using two numerical integration rules: Riemann rule (a first-order method) and Simpson rule (a fourth-order method).
The error $\delta$ is the $L_2$-norm of the difference between the output electric field $E$ of the propagator and $F_{\text{AD}}$, as in
\begin{equation}
    \delta = \|E - F_{\text{AD}}\|_2.
\end{equation}
\begin{figure}[t]
    \centering
    \includegraphics[width=0.5\columnwidth]{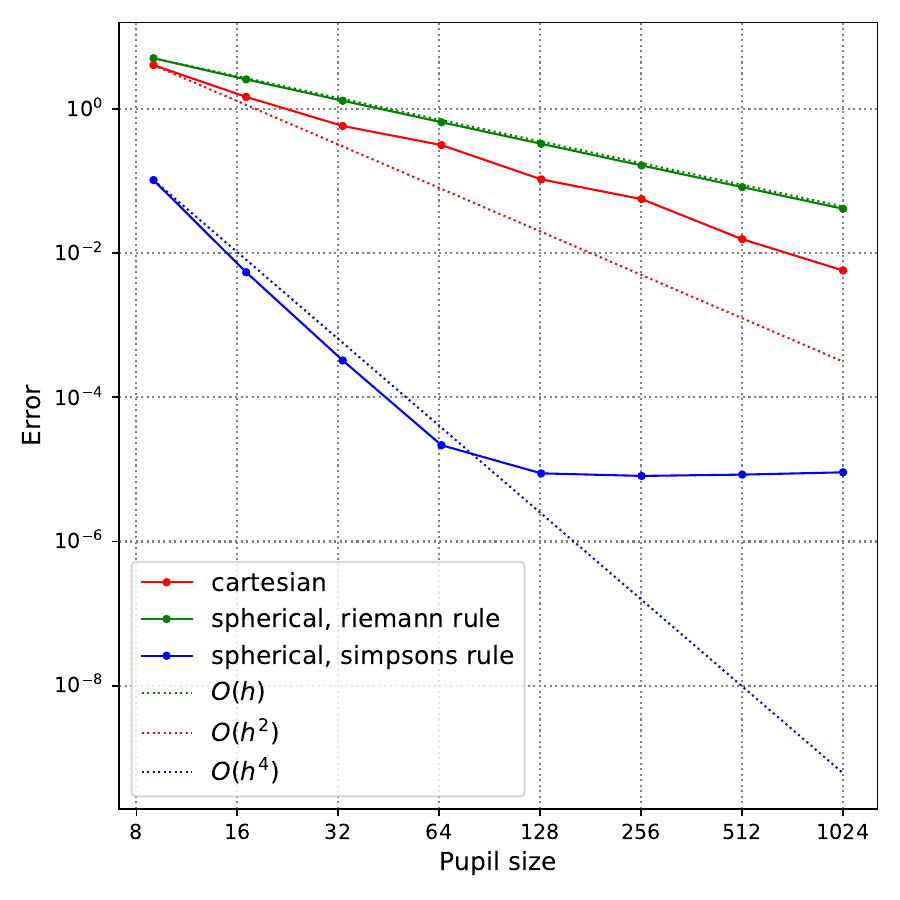}
    \caption{Accuracy of the Cartesian (red) and the spherical propagators using Riemann rule (green) and Simpson rule (blue).
    The step size of integration is $h$.
    \label{fig:benchmark-accuracy}}
\end{figure}

We observe in Figure \ref{fig:benchmark-accuracy} that the error decreases as the number of points in the integration domain increases in all cases.
The spherical propagator has a linear convergence rate using Riemann rule (green) and 4th-order convergence rate using Simpson rule (blue), which correlates well with their expected accuracy.
The Cartesian propagator, based on our custom FFT, shows a convergence rate between first- and second-order.




\section{Conclusion}

In this work, we have introduced a unifying theory for accurate PSF models, revisiting previous approaches and generalizing correction factors. This framework demonstrates the equivalence of Cartesian and spherical methods as different parametrizations of the same propagation integral, providing a simplified understanding of light propagation in high-NA systems. 

We have also developed a high-performance implementation in PyTorch, which is open-source and supported by comprehensive documentation. This library allows for efficient CPU/GPU computation of PSFs, with functionalities such as automatic differentiation that make it particularly suitable for optimization pipelines. 

In practice, the choice of propagators depends on the symmetry of the pupil function. 
Based on our benchmark, for axisymmetric pupil functions, spherical propagators are recommended thanks to their high accuracy and scalability. They are also particularly amenable to GPU parallelization. 
For non-axisymmetric cases, such as those involving Zernike aberrations or specialized phase masks, Cartesian propagators are required. 
In all these cases, the difference in computational time is relatively modest for pupil sizes up to a few hundred pixels. 
Hence, the vectorial Cartesian propagator could be a solid default choice in most applications. 

We hope for our PyTorch library to contribute to the rapidly growing field of applying deep neural networks on physical imaging models \cite{belthangady2019applications}. 
Learning-based methods have demonstrated their effectiveness in various applications, such as deconvolution \cite{weigert2018content, li2022incorporating, yanny2022deep} and 3D SMLM \cite{sage2019super}, with state-of-the-art deep-learning tools such as DeepSTORM \cite{nehme2020deepstorm3d} and DECODE \cite{speiser2021deep}.
For instance, some use cases of our framework include generating large reference datasets to train networks or adapting images based on physical rules to enable learning in self-supervised inversions \cite{kobayashi2020image} and generative adversarial networks \cite{cachia2023fluorescence}. 

\section*{Funding}
Vasiliki Stergiopoulou acknowledges funding from the Swiss National Science Foundation under Grant CRSII5\_213521, “DigiLight - Programmable Third-Harmonic Generation (THG) Microscopy Applied to Advanced Manufacturing.”
Jonathan Dong acknowledges funding from the Swiss National Science Foundation (Grant PZ00P2\_216211). 

\section*{Acknowledgment}
We would like to thank Eric Sinner for his valuable help in software engineering.

\bibliographystyle{ieeetr}
\bibliography{refs.bib}

\begin{thebibliography}{10}

\bibitem{matthews2010metroloj}
C.~Matthews and F.~P. Cordelieres, ``{MetroloJ}: An {ImageJ} plugin to help monitor microscopes’ health,'' in {\em ImageJ User \& Developer Conference Proceedings}, pp.~1--6, 2010.

\bibitem{theer2014psfj}
P.~Theer, C.~Mongis, and M.~Knop, ``{PSFj}: Know your fluorescence microscope,'' {\em Nature Methods}, vol.~11, no.~10, pp.~981--982, 2014.

\bibitem{miora2024calculating}
R.~H.~D. Miora, E.~Rohwer, M.~Kielhorn, C.~Sheppard, G.~Bosman, and R.~Heintzmann, ``Calculating point spread functions: Methods, pitfalls, and solutions,'' {\em Optics Express}, vol.~32, no.~16, pp.~27278--27302, 2024.

\bibitem{nelson2022monitoring}
G.~Nelson, A.~Payne-Dwyer, and {QUAREP-LiMi Working Group 5}, ``Monitoring the point spread function for quality control of confocal microscopes,'' Oct. 2022.

\bibitem{lelek2021single}
M.~Lelek, M.~T. Gyparaki, G.~Beliu, F.~Schueder, J.~Griffi{\'e}, S.~Manley, R.~Jungmann, M.~Sauer, M.~Lakadamyali, and C.~Zimmer, ``Single-molecule localization microscopy,'' {\em Nature reviews Methods Primers}, vol.~1, no.~1, p.~39, 2021.

\bibitem{sibarita2005deconvolution}
J.-B. Sibarita, ``Deconvolution microscopy,'' {\em Microscopy Techniques}, vol.~95, pp.~201--243, 2005.

\bibitem{sage2017deconvolutionlab2}
D.~Sage, L.~Donati, F.~Soulez, D.~Fortun, G.~Schmit, A.~Seitz, R.~Guiet, C.~Vonesch, and M.~Unser, ``{DeconvolutionLab2}: An open-source software for deconvolution microscopy,'' {\em Methods}, vol.~115, pp.~28--41, 2017.

\bibitem{heintzmann1999laterally}
R.~Heintzmann and C.~G. Cremer, ``Laterally modulated excitation microscopy: Improvement of resolution by using a diffraction grating,'' in {\em Optical Biopsies and Microscopic Techniques III}, vol.~3568, pp.~185--196, SPIE, 1999.

\bibitem{gustafsson2000surpassing}
M.~G. Gustafsson, ``Surpassing the lateral resolution limit by a factor of two using structured illumination microscopy,'' {\em Journal of Microscopy}, vol.~198, no.~2, pp.~82--87, 2000.

\bibitem{dertinger2009fast}
T.~Dertinger, R.~Colyer, G.~Iyer, S.~Weiss, and J.~Enderlein, ``Fast, background-free, {3D} super-resolution optical fluctuation imaging (sofi),'' {\em Proceedings of the National Academy of Sciences}, vol.~106, no.~52, pp.~22287--22292, 2009.

\bibitem{stergiopoulou_BioIm}
V.~Stergiopoulou, L.~Calatroni, H.~de~Morais~Goulart, S.~Schaub, and L.~Blanc-Féraud, ``{COL0RME}: Super-resolution microscopy based on sparse blinking/fluctuating fluorophore localization and intensity estimation,'' {\em Biological Imaging}, vol.~2, 2022.

\bibitem{hell2007far}
S.~W. Hell, ``Far-field optical nanoscopy,'' {\em Science}, vol.~316, no.~5828, pp.~1153--1158, 2007.

\bibitem{balzarotti2017nanometer}
F.~Balzarotti, Y.~Eilers, K.~C. Gwosch, A.~H. Gynn{\aa}, V.~Westphal, F.~D. Stefani, J.~Elf, and S.~W. Hell, ``Nanometer resolution imaging and tracking of fluorescent molecules with minimal photon fluxes,'' {\em Science}, vol.~355, no.~6325, pp.~606--612, 2017.

\bibitem{marin2021pymevisualize}
Z.~Marin, M.~Graff, A.~E. Barentine, C.~Soeller, K.~K.~H. Chung, L.~A. Fuentes, and D.~Baddeley, ``{PYMEVisualize}: An open-source tool for exploring {3D} super-resolution data,'' {\em Nature Methods}, vol.~18, no.~6, pp.~582--584, 2021.

\bibitem{richards1959electromagnetic}
B.~Richards and E.~Wolf, ``Electromagnetic diffraction in optical systems, {II.} {S}tructure of the image field in an aplanatic system,'' {\em Proceedings of the Royal Society of London. Series A. Mathematical and Physical Sciences}, vol.~253, no.~1274, pp.~358--379, 1959.

\bibitem{leutenegger2006fast}
M.~Leutenegger, R.~Rao, R.~A. Leitgeb, and T.~Lasser, ``Fast focus field calculations,'' {\em Optics Express}, vol.~14, no.~23, pp.~11277--11291, 2006.

\bibitem{aguet2009super}
F.~Aguet, {\em Super-resolution fluorescence microscopy based on physical models}.
\newblock PhD thesis, EPFL, 2009.

\bibitem{Novotny_Hecht_2012}
L.~Novotny and B.~Hecht, {\em Principles of Nano-Optics}.
\newblock Cambridge: Cambridge University Press, 2~ed., 2012.

\bibitem{gibson1991experimental}
S.~F. Gibson and F.~Lanni, ``Experimental test of an analytical model of aberration in an oil-immersion objective lens used in three-dimensional light microscopy,'' {\em Journal of the Optical Society of America A}, vol.~8, no.~10, pp.~1601--1613, 1991.

\bibitem{kirshner20133}
H.~Kirshner, A.~Fran{\c{c}}ois, D.~Sage, and M.~Unser, ``{3-D} {PSF} fitting for fluorescence microscopy: Implementation and localization application,'' {\em Journal of Microscopy}, vol.~249, no.~1, pp.~13--25, 2013.

\bibitem{nasse2010realistic}
M.~J. Nasse and J.~C. Woehl, ``Realistic modeling of the illumination point spread function in confocal scanning optical microscopy,'' {\em Josa a}, vol.~27, no.~2, pp.~295--302, 2010.

\bibitem{schneider2024interactive}
M.~C. Schneider, F.~Hinterer, A.~Jesacher, and G.~J. Sch{\"u}tz, ``Interactive simulation and visualization of point spread functions in single molecule imaging,'' {\em Optics Communications}, vol.~560, p.~130463, 2024.

\bibitem{caprile2022pyfocus}
F.~Caprile, L.~A. Masullo, and F.~D. Stefani, ``Pyfocus--a {Python} package for vectorial calculations of focused optical fields under realistic conditions. {Application} to toroidal foci,'' {\em Computer Physics Communications}, vol.~275, p.~108315, 2022.

\bibitem{prigent2023spitfir}
S.~Prigent, H.-N. Nguyen, L.~Leconte, C.~A. Valades-Cruz, B.~Hajj, J.~Salamero, and C.~Kervrann, ``{SPITFIR} (e): A supermaneuverable algorithm for fast denoising and deconvolution of {3D} fluorescence microscopy images and videos,'' {\em Scientific Reports}, vol.~13, no.~1, p.~1489, 2023.

\bibitem{li2018real}
Y.~Li, M.~Mund, P.~Hoess, J.~Deschamps, U.~Matti, B.~Nijmeijer, V.~J. Sabinina, J.~Ellenberg, I.~Schoen, and J.~Ries, ``Real-time {3D} single-molecule localization using experimental point spread functions,'' {\em Nature Methods}, vol.~15, no.~5, pp.~367--369, 2018.

\bibitem{dong2021fundamental}
J.~Dong, D.~Maestre, C.~Conrad-Billroth, and T.~Juffmann, ``Fundamental bounds on the precision of {iSCAT}, {COBRI} and dark-field microscopy for {3D} localization and mass photometry,'' {\em Journal of Physics D: Applied Physics}, vol.~54, no.~39, p.~394002, 2021.

\bibitem{shechtman2014optimal}
Y.~Shechtman, S.~J. Sahl, A.~S. Backer, and W.~E. Moerner, ``Optimal point spread function design for {3D} imaging,'' {\em Physical Review Letters}, vol.~113, no.~13, p.~133902, 2014.

\bibitem{opatovski2024depth}
N.~Opatovski, E.~Nehme, N.~Zoref, I.~Barzilai, R.~Orange~Kedem, B.~Ferdman, P.~Keselman, O.~Alalouf, and Y.~Shechtman, ``Depth-enhanced high-throughput microscopy by compact {PSF} engineering,'' {\em Nature Communications}, vol.~15, no.~1, p.~4861, 2024.

\bibitem{Liu:25}
Y.~Liu, J.~Dong, J.~A. Maya, F.~Balzarotti, and M.~Unser, ``Point-spread-function engineering in {MINFLUX}: ptimality of donut and half-moon excitation patterns,'' {\em Optics Letters}, vol.~50, pp.~37--40, Jan 2025.

\bibitem{sinko2014teststorm}
J.~Sink{\'o}, R.~K{\'a}konyi, E.~Rees, D.~Metcalf, A.~E. Knight, C.~F. Kaminski, G.~Szab{\'o}, and M.~Erd{\'e}lyi, ``{TestSTORM}: Simulator for optimizing sample labeling and image acquisition in localization based super-resolution microscopy,'' {\em Biomedical Optics Express}, vol.~5, no.~3, pp.~778--787, 2014.

\bibitem{sage2015quantitative}
D.~Sage, H.~Kirshner, T.~Pengo, N.~Stuurman, J.~Min, S.~Manley, and M.~Unser, ``Quantitative evaluation of software packages for single-molecule localization microscopy,'' {\em Nature Methods}, vol.~12, no.~8, pp.~717--724, 2015.

\bibitem{wieser2008tracking}
S.~Wieser and G.~J. Sch{\"u}tz, ``Tracking single molecules in the live cell plasma membrane—{D}o’s and {D}on’t’s,'' {\em Methods}, vol.~46, no.~2, pp.~131--140, 2008.

\bibitem{shen2017single}
H.~Shen, L.~J. Tauzin, R.~Baiyasi, W.~Wang, N.~Moringo, B.~Shuang, and C.~F. Landes, ``Single particle tracking: From theory to biophysical applications,'' {\em Chemical Reviews}, vol.~117, no.~11, pp.~7331--7376, 2017.

\bibitem{griffie2020virtual}
J.~Griffi{\'e}, T.~Pham, C.~Sieben, R.~Lang, V.~Cevher, S.~Holden, M.~Unser, S.~Manley, and D.~Sage, ``Virtual-{SMLM}, a virtual environment for real-time interactive {SMLM} acquisition,'' {\em bioRxiv}, 03 2020.

\bibitem{bourgeois2023single}
D.~Bourgeois, ``Single molecule imaging simulations with advanced fluorophore photophysics,'' {\em Communications Biology}, vol.~6, no.~1, p.~53, 2023.

\bibitem{torok1997electromagnetic}
P.~T{\"o}r{\"o}k and P.~Varga, ``Electromagnetic diffraction of light focused through a stratified medium,'' {\em Applied Optics}, vol.~36, no.~11, pp.~2305--2312, 1997.

\bibitem{napari}
N.~Sofroniew, T.~Lambert, G.~Bokota, J.~Nunez-Iglesias, P.~Sobolewski, A.~Sweet, L.~Gaifas, K.~Evans, A.~Burt, D.~Doncila~Pop, K.~Yamauchi, M.~Weber~Mendonça, G.~Buckley, W.-M. Vierdag, L.~Royer, A.~Can~Solak, K.~I.~S. Harrington, J.~Ahlers, D.~Althviz~Moré, O.~Amsalem, A.~Anderson, A.~Annex, P.~Boone, J.~Bragantini, M.~Bussonnier, C.~Caporal, J.~Eglinger, A.~Eisenbarth, J.~Freeman, C.~Gohlke, K.~Gunalan, H.~Har-Gil, M.~Harfouche, V.~Hilsenstein, K.~Hutchings, J.~Lauer, G.~Lichtner, Z.~Liu, L.~Liu, A.~Lowe, L.~Marconato, S.~Martin, A.~McGovern, L.~Migas, N.~Miller, H.~Muñoz, J.-H. Müller, C.~Nauroth-Kreß, D.~Palecek, C.~Pape, E.~Perlman, K.~Pevey, G.~Peña-Castellanos, A.~Pierré, D.~Pinto, J.~Rodríguez-Guerra, D.~Ross, C.~T. Russell, J.~Ryan, G.~Selzer, M.~Smith, P.~Smith, K.~Sofiiuk, J.~Soltwedel, D.~Stansby, J.~Vanaret, P.~Wadhwa, M.~Weigert, J.~Windhager, P.~Winston, and R.~Zhao, ``napari: a multi-dimensional image viewer for python,'' Sept. 2024.

\bibitem{deb2025chromatix}
D.~Deb, G.-J. Both, E.~Bezzam, A.~Kohli, S.~Yang, A.~Chaware, C.~Allier, C.~Cai, G.~Anderberg, M.~H. Eybposh, {\em et~al.}, ``Chromatix: a differentiable, gpu-accelerated wave-optics library,'' {\em bioRxiv}, pp.~2025--04, 2025.

\bibitem{debye1908lichtdruck}
P.~J.~W. Debye, {\em Der lichtdruck auf kugeln von beliebigem material}.
\newblock PhD thesis, Ludwig-Maximilians Universit{\"a}t M{\"u}nchen, 1908.

\bibitem{wolf1981conditions}
E.~Wolf and Y.~Li, ``Conditions for the validity of the {D}ebye integral representation of focused fields,'' {\em Optics Communications}, vol.~39, no.~4, pp.~205--210, 1981.

\bibitem{egner1999equivalence}
A.~Egner and S.~Hell, ``Equivalence of the {Huygens}--{Fresnel} and {Debye} approach for the calculation of high aperture point-spread functions in the presence of refractive index mismatch,'' {\em Journal of Microscopy}, vol.~193, no.~3, pp.~244--249, 1999.

\bibitem{torok1995electromagnetic}
P.~T{\"o}r{\"o}k, P.~Varga, Z.~Laczik, and G.~Booker, ``Electromagnetic diffraction of light focused through a planar interface between materials of mismatched refractive indices: {A}n integral representation,'' {\em Journal of the Optical Society of America A}, vol.~12, no.~2, pp.~325--332, 1995.

\bibitem{goodman2005introduction}
J.~Goodman, {\em Introduction to Fourier Optics}.
\newblock McGraw-Hill physical and quantum electronics series, W. H. Freeman, 2005.

\bibitem{liu2024universal}
S.~Liu, J.~Chen, J.~Hellgoth, L.-R. M{\"u}ller, B.~Ferdman, C.~Karras, D.~Xiao, K.~A. Lidke, R.~Heintzmann, Y.~Shechtman, {\em et~al.}, ``Universal inverse modeling of point spread functions for {SMLM} localization and microscope characterization,'' {\em Nature Methods}, vol.~21, no.~6, pp.~1082--1093, 2024.

\bibitem{stoer_introduction_2002}
J.~Stoer and R.~Bulirsch, {\em Introduction to {Numerical} {Analysis}}, vol.~12 of {\em Texts in {Applied} {Mathematics}}.
\newblock New York, NY: Springer New York, 2002.

\bibitem{barker2022introducing}
M.~Barker, N.~P. Chue~Hong, D.~S. Katz, A.-L. Lamprecht, C.~Martinez-Ortiz, F.~Psomopoulos, J.~Harrow, L.~J. Castro, M.~Gruenpeter, P.~A. Martinez, {\em et~al.}, ``Introducing the fair principles for research software,'' {\em Scientific Data}, vol.~9, no.~1, p.~622, 2022.

\bibitem{kao1994tracking}
H.~P. Kao and A.~Verkman, ``Tracking of single fluorescent particles in three dimensions: Use of cylindrical optics to encode particle position,'' {\em Biophysical Journal}, vol.~67, no.~3, pp.~1291--1300, 1994.

\bibitem{belthangady2019applications}
C.~Belthangady and L.~A. Royer, ``Applications, promises, and pitfalls of deep learning for fluorescence image reconstruction,'' {\em Nature Methods}, vol.~16, no.~12, pp.~1215--1225, 2019.

\bibitem{weigert2018content}
M.~Weigert, U.~Schmidt, T.~Boothe, A.~M{\"u}ller, A.~Dibrov, A.~Jain, B.~Wilhelm, D.~Schmidt, C.~Broaddus, S.~Culley, {\em et~al.}, ``Content-aware image restoration: pushing the limits of fluorescence microscopy,'' {\em Nature Methods}, vol.~15, no.~12, pp.~1090--1097, 2018.

\bibitem{li2022incorporating}
Y.~Li, Y.~Su, M.~Guo, X.~Han, J.~Liu, H.~D. Vishwasrao, X.~Li, R.~Christensen, T.~Sengupta, M.~W. Moyle, {\em et~al.}, ``Incorporating the image formation process into deep learning improves network performance,'' {\em Nature Methods}, vol.~19, no.~11, pp.~1427--1437, 2022.

\bibitem{yanny2022deep}
K.~Yanny, K.~Monakhova, R.~W. Shuai, and L.~Waller, ``Deep learning for fast spatially varying deconvolution,'' {\em Optica}, vol.~9, no.~1, pp.~96--99, 2022.

\bibitem{sage2019super}
D.~Sage, T.-A. Pham, H.~Babcock, T.~Lukes, T.~Pengo, J.~Chao, R.~Velmurugan, A.~Herbert, A.~Agrawal, S.~Colabrese, {\em et~al.}, ``Super-resolution fight club: assessment of {2D} and {3D} single-molecule localization microscopy software,'' {\em Nature Methods}, vol.~16, no.~5, pp.~387--395, 2019.

\bibitem{nehme2020deepstorm3d}
E.~Nehme, D.~Freedman, R.~Gordon, B.~Ferdman, L.~E. Weiss, O.~Alalouf, T.~Naor, R.~Orange, T.~Michaeli, and Y.~Shechtman, ``{DeepSTORM3D}: dense {3D} localization microscopy and {PSF} design by deep learning,'' {\em Nature Methods}, vol.~17, no.~7, pp.~734--740, 2020.

\bibitem{speiser2021deep}
A.~Speiser, L.-R. M{\"u}ller, P.~Hoess, U.~Matti, C.~J. Obara, W.~R. Legant, A.~Kreshuk, J.~H. Macke, J.~Ries, and S.~C. Turaga, ``Deep learning enables fast and dense single-molecule localization with high accuracy,'' {\em Nature Methods}, vol.~18, no.~9, pp.~1082--1090, 2021.

\bibitem{kobayashi2020image}
H.~Kobayashi, A.~C. Solak, J.~Batson, and L.~A. Royer, ``Image deconvolution via noise-tolerant self-supervised inversion,'' {\em arXiv preprint arXiv:2006.06156}, 2020.

\bibitem{cachia2023fluorescence}
M.~Cachia, V.~Stergiopoulou, L.~Calatroni, S.~Schaub, and L.~Blanc-F{\'e}raud, ``Fluorescence image deconvolution microscopy via generative adversarial learning ({FluoGAN}),'' {\em Inverse Problems}, vol.~39, no.~5, p.~054006, 2023.

\end{thebibliography}

\end{document}